\documentstyle[aps,preprint]{revtex}
\begin{document}

\draft

\title{Direct test of composite fermion model in quantum
	 Hall systems}
\author{Sudhansu S. Mandal and V. Ravishankar}
\address{Department of Physics, Indian Institute of technology,
Kanpur -- 208 016, INDIA }

\maketitle

\begin{abstract}

We show that neutron scattering and Raman
scattering experiments can unambiguously determine a
composite fermion parameter, viz., the effective number of
Landau Levels filled by the composite fermions. For this purpose, one
needs partially polarized or more preferably unpolarized quantum Hall
states. We further find that spin correlation function acts as
an order parameter in the spin transition.

\end{abstract}


\section{INTRODUCTION}

In recent years, fractional quantum Hall effect (FQHE) which is
believed to arise due to complicated electron electron
interactions in the presence of high magnetic field $(B)$
(perpendicular to the plane of two dimensional electron system)
has drawn much interest of physicists. The composite fermion model
(CFM), which is proposed by Jain \cite{jain}, is by now fairly
well established in these systems. In this model the interaction
of one electron with all the others is replaced by attaching an even
number $(2s)$ of flux quanta ( in the units of $2\pi /e$)
to each electron. In
the mean field (MF) approximation, these fluxes produce a uniform
magnetic field such that the effective Landau levels (LL)
formed by the effective magnetic field $\bar{B}=B-(2\pi /e)\rho
(2s)$, (where $\rho$ is the mean particle density),
can accommodate the particles in an integral number $(p)$
of effective LL. The integer quantum Hall effect (IQHE) at
integer filling $p$ by these composite fermions (CF) leads to FQHE
at the filling fraction $\nu =p/(2sp +1)$
in the original electronic system.
Later Lopez and Fradkin \cite{lopez1}
have developed a formalism to study FQHE within this model by
the introduction of an appropriate Chern-Simons (CS) gauge field.

Successful though the model is,
Laughlin \cite{laugh1} has criticised the model on the grounds
that it does not
make any reference either to fractionally charged quasi-particles
\cite{laugh2}
or their fractional statistics \cite{halp1}
with which one could construct the hierarchial FQHE
states by their condensations \cite{halp1,hald}.
In this hierarchial picture,
the elementary excitations in the state with
filling fraction $\nu = p/(2p+1)$ have
charge $\pm e/(2p+1)$ \cite{laugh2,halp1,hald}.
Notice that on the other hand, in the CFM,
quasi-particles have charge $-e$
with $2s$ vortices \cite{fnote1}.

On the other hand, good experimental evidence
for the existence of CF in FQHE systems has
emerged recently \cite{du1,will,kang,lead,du2,mano,gold}.
As Halperin, Lee and Read (HLR) \cite{hlr} emphasized,
the single particle excitation gap of CF,
corresponding to the state with filling fraction
$\nu =p/(2sp + 1)$, is the effective cyclotron
frequency $\bar{\omega}_c$
which is determined by the effective field $\bar{B}$.
Du et al \cite{du1} find that their results on the activation of
the diagonal resistivity $\rho_{xx}$ is consistent with the
above interpretation. More significantly, CFM makes the
remarkable prediction that at $\nu =1/2s$, the effective field
$\bar{B} =0$. The properties of the half-filled LL have been
studied extensively by HLR \cite{hlr} employing the CF picture.
Indeed at $\nu =1/2s$, the CF should have a well
defined Fermi surface which has been verified experimentally by
Willett et al \cite{will} and Kang et al \cite{kang} by
observing cyclotron motion of CF near $\nu =1/2$.
Three recent experiments \cite{lead,du2,mano}
have treated the oscillations
in $\rho_{xx}$ around $\nu =1/2$ as Shubnikosov-de
Haas oscillations (SDHO) of CF, in analogy to SDHO of free electrons
near $B=0$. However, Leadley et al \cite{lead} have reported a
finite effective mass $m^\ast$ of CF at $\bar{B}=0$, and
$m^\ast$ increases linearly with $\vert \bar{B}\vert $,
while Du et al \cite{du2}
and Manoharan et al \cite{mano} have observed `drastic
enhancement' of CF mass as $\nu \rightarrow  1/2$, indicating a
novel Fermi liquid at $\nu =1/2$.
Goldman et al \cite{gold} have reported the confirmation of
the existence of CF by observing negatively charged carriers
to form a Fermi sea near $\nu =1/2$ in a magnetic focussing
experiment, and they have also found that the charge carriers
experience an effective magnetic field $\bar{B}$. The main
conclusion of the above experiments is that the dynamics of the
charged particles is governed by $\bar{B}$, rather than the
applied magnetic field $B$.

All the above experiments which are strongly in favour of the
existence of CF are still rather incomplete in the
sense that none of them determine either of the composite
fermion parameters, viz, the effective number of LL $(p)$ or
the number of flux quanta $(2s)$ attached to each electron directly.
The gap measurements do not determine $p$ unambiguously as the
parametrization of activation of $\rho_{xx}$ is not unique and
$m^\ast$ also changes with $\bar{B}$. Here we propose
experiments which would determine $\vert p\vert$ unambiguously.
The other parameter $2s$ can be found
out from the knowledge of filling
fraction $\nu =\vert p\vert /(2s\vert p \vert \pm 1)$. To that
end, we need the FQHE states which are either partially polarized
or unpolarized. Indeed, they are central to our analysis because the
wave functions for fully polarized quantum Hall states (QHS) depend
solely on $\nu$ \cite{lopez3}, while on the other
hand, as we have shown recently \cite{wfn},
the wave functions for unpolarized or partially polarized
QHS depend on any two parameters among $p, s$, and $\nu$.

We compute here the spin density correlation (SDC) in QHS and find that
it is, in fact, an order parameter in the spin transitions from
spin unpolarized or partially polarized phases to their fully
polarized phase. In fact, the ratio of SDC for unpolarized and
fully polarized phase would determine the effective number
of LL which are filled. The static charge density structure
factor depends on $p$ only in the unpolarized phase. These are
independent of $m^\ast $. Therefore experiments like neutron
scattering would determine $p$ avoiding any complexity arising
from the dependence of effective mass on $\bar{B}$.
We also determine the collective
excitations from the poles of charge density correlations (CDC)
and SDC.
For the former, there is no mode near $\bar{\omega}_c$
but near the actual cyclotron frequency $\omega_c$,
irrespective of the spin phase. On the other hand, SDC
is shown to possess an undispersed pole {\em exactly}
at $\bar{\omega}_c$
in unpolarized and partially polarized phases. Therefore,
depolarized Raman scattering would again
determine the exact value of
$\bar{\omega}_c$ which is also a measure of $p$.

\section{BRIEF REVIEW}

Recently we have developed an abelian doublet model \cite{model}
employing a doublet of CS gauge fields, by which we can account
for all the known filling fractions with different possible spin
polarizations. Further, we
have extracted the wave functions \cite{wfn} as well from the
correlations for arbitrarily polarized QHS. We, therefore, do
not repeat the details of either the model or the computation of
correlation functions, but present only the essential features.

\subsection{The Model}

To describe in brief,
consider a two-dimensional system of
spin-1/2 interacting electrons in the presence
of uniform magnetic field
perpendicular to the plane.
The complicated interaction among electrons is represented by the
interaction of electrons with CS gauge fields
and weak (short-ranged) fermion-fermion interaction as we
discuss below. We consider quantum Hall effect
in low but non-zero Zeeman
energy limit.
The dynamics of the system is represented by the Lagrangian
density
\begin{eqnarray}
{\cal L} &=& \psi_\uparrow^\dagger {\cal D}
     (A_\mu^\uparrow +a_\mu^\uparrow )\psi_\uparrow
   + \psi_\downarrow^\dagger {\cal D}(A_\mu^\downarrow +
   a_\mu^\downarrow )\psi_\downarrow
   +{1 \over 2} \tilde{a}_\mu\epsilon^{\mu\nu \lambda } \Theta
   \partial_\nu a_\lambda \nonumber \\
   & & -eA_0^{\rm in}\rho +{1 \over 2}\int d^3 x^\prime A_0^{\rm
   in}(x)V^{-1}(x-x^\prime)A_0^{\rm in} (x^\prime)\; .
\label{eq1}
\end{eqnarray}
Here $\psi $ is the fermionic field and $\uparrow (\downarrow )$
represents spin-up (down),
\begin{equation}
{\cal D}(A_\mu^r +a_\mu^r) = iD_0^r +(1/2m^\ast)
D_k^{r\, 2} +\mu +(g/2)\mu_B (B+B^r+b^r)\sigma \; ,
\label{eq2}
\end{equation}
with $D_\mu^r =\partial_\mu -ie
(A_\mu +A_\mu^r+a_\mu^r)$ where $A_\mu $
is the external electro-magnetic field which interacts with all
the electrons while $A_\mu^r$ and $a_\mu^r$ are the external
probe \cite{wfn} and the CS gauge field
respectively, interacting with {\it only} the particles having
spin indices $r= \uparrow \, ,\, \downarrow $. The field
$A_0^{\rm in}$
is identified as an internal scalar potential.
Fixed mean particle density
$\rho$ is represented by the chemical potential
$\mu$ which acts as a Lagrange multiplier.
Note that the Zeeman term includes all the three kinds
of magnetic fields. $\mu_B$ is the Bohr-magneton, and $\sigma =+1
(-1)$ for spin-up (down) electrons. We have introduced
an abelian doublet of CS gauge fields in (\ref{eq1}) as
\begin{equation}
a_\mu = \left( \begin{array}{c}
    a_\mu^\uparrow \\
    a_\mu^\downarrow \end{array}
    \right) \; ,
\label{eq3}
\end{equation}
and the strength of the real symmetric matrix valued
CS parameter is taken to be
\begin{equation}
\Theta = \left( \begin{array}{cc}
    \theta_1 & \theta_2 \\
    \theta_2 & \theta_1 \end{array}
    \right)  \; .
\label{eq4}
\end{equation}
$\tilde{a}_\mu $ is the transpose of the doublet field $a_\mu$.
The fourth term in Eq.~(\ref{eq1}) describes the charge
neutrality of the system. Finally, $V^{-1}(x-x^\prime )$
is the inverse
of the electron interaction potential (in the operator sense).
The usual fermion interaction term in quartic form would be
achieved by an integration over $A_0^{\rm in}$ field.
The values of $\theta_1$ and $\theta_2$ must be consistent
with the composite fermion requirement.

We then diagonalize the matrix $\Theta $, with the eigen values
$ \theta_\pm = \theta_1 \pm \theta_2$.
In the eigen basis,
by simple rescalings, Eq.~(\ref{eq1}) may be written as
\begin{eqnarray}
{\cal L} &=& \psi_\uparrow^\dagger {\cal D}
(A_\mu^\uparrow +a_\mu^+ +a_\mu^-)\psi_\uparrow
  + \psi_\downarrow^\dagger {\cal D} (A_\mu^\downarrow +
  a_\mu^+ -a_\mu^-)\psi_\downarrow
  +{\theta_+ \over 2 }\epsilon^{\mu\nu \lambda}
  a_\mu^+\partial_\nu a_\lambda^+ \nonumber \\
 & & +{\theta_- \over 2 }\epsilon^{\mu\nu \lambda}
  a_\mu^-\partial_\nu a_\lambda^-
    -eA_0^{\rm in}\rho +{1 \over 2}\int d^3 x^\prime A_0^{\rm
   in}(x)V^{-1}(x-x^\prime)A_0^{\rm in} (x^\prime)\; .
 \label{eq5}
\end{eqnarray}
This incorporates the idea that each electron, in general, has
two kinds of vortices associated with it --- while they
interact in phase with spin up particles, spin down particles
get their out of phase contributions.

Consider the case $ \theta_- =0$. Here,
$a_\mu^-$ decouples dynamically and merely plays the role
of a Lagrange multiplier: $( \partial {\cal L} / \partial
a_0^- )= \rho_ \uparrow - \rho_ \downarrow \equiv 0$, where
$\rho_ \uparrow (\rho_ \downarrow )$ is the density for spin-up
(down) particles.
We then parametrize $ \theta = (e^2 /2\pi)(1/2s)$ ($s$
is an integer) in order to impose the composite fermion picture
-- fermions are attached with $2s$ vortices. In the mean field
(MF) ansatz, these vortices produce an average CS magnetic field
$\langle b^+ \rangle = -e\rho/\theta_+ $. These choice of the
parameters lead to the unpolarized QHS.

On the other hand, for obtaining partially polarized QHS, we
parametrize $\theta_\pm = (e^2/2\pi)(1/s_\pm )$ and set $s_+
=2s$ and $s_- =0$. In this case, the field $a_\mu^-$ provides a
vanishing mean magnetic field $\langle b^- \rangle$, and does
not contribute to tree level (in contrast to the unpolarized
case where $a_\mu^-$ is completely nondynamical). Composite
fermion picture is enforced by the choice of $s_+ =2s$. Thus in
the MF ansatz, CS magnetic field produced by the particles is
$\langle b^+ \rangle =-e\rho /\theta_+ $.

In both the above cases, mean magnetic field for all the
particles, irrespective of their spin, is given by
$\bar{B}=B+\langle b^+ \rangle $.
Let $p_ \uparrow (p_ \downarrow )$
be the number of effective Landau levels (LL)
formed by $\bar{B}^+$ filled by spin up
(down) particles. This leads to the actual filling fraction
and the spin density to be
\begin{equation}
\nu = {p_ \uparrow + p_ \downarrow \over 2s(p_ \uparrow + p_
\downarrow ) +1}\;\; ; \; \;
\Delta\rho = \rho \left( {p_ \uparrow -p_ \downarrow
\over  p_ \uparrow + p_ \downarrow } \right) \; .
\label{eq6}
\end{equation}
Note that $p_ \uparrow $ and
$p_ \downarrow $ can be negative integers as well in which case
$\bar{B}^+$ is antiparallel to $B$.
The effective cyclotron frequency $\bar{\omega}_c =
e\bar{B}/m^\ast$
is related to $\omega_c = eB/m^\ast $ by
$\omega_c =\bar{\omega}_c [2s(p_\uparrow +p_
\downarrow ) +1 ]$.
For unpolarized QHS, $p_\uparrow = p_\downarrow =p $ (say) and
therefore the states with filling fraction $\nu =2p/(4sp+1)$ are
spin unpolarized in the limit of small Zeeman energy. In this
limit, $p_ \uparrow = p_\downarrow +1 $ for partially polarized
states with $\nu = (p_\uparrow +p_\downarrow )/
(2s(p_\uparrow +p_\downarrow  )+1)$ and
$\Delta \rho / \rho = 1/(p_\uparrow +p_\downarrow )$.
Fully polarized Laughlin
states are obtained for $p_ \uparrow =1 \, , \, p_\downarrow =0$.

\subsection{Effective Action}

Employing the above MF ansatz, we then evaluate one-loop
effective action for the gauge
fields to be
\begin{eqnarray}
S_{\rm eff} &=&
 -{1 \over 2}\int {d^3 q \over (2\pi)^3} (A_\mu^\uparrow
 +a_\mu^+ +a_\mu^- )\Pi^{\mu\nu}_ \uparrow (\omega \, , \, {\bf
 q}^2 ) (A_\nu^\uparrow +a_\mu^+ +a_\mu^- ) \nonumber \\ & &
 -{1 \over 2}\int {d^3 q \over (2\pi)^3} (A_\mu^\downarrow
 +a_\mu^+ -a_\mu^- )\Pi^{\mu\nu}_ \downarrow (\omega \, , \, {\bf
 q}^2 ) (A_\nu^\downarrow +a_\mu^+ -a_\mu^- ) \nonumber \\
 & & + {i \over 2} \int {d^3 q \over (2\pi)^3} \left[
 {\theta_+ \over 2}\epsilon^{\mu\nu\lambda}a_\mu^+ q_\nu
 a_\lambda^+
 +{\theta_- \over 2}\epsilon^{\mu\nu\lambda}a_\mu^- q_\nu
 a_\lambda^- \right] \nonumber \\ & &
 +{1\over 2}\int {d^3 q \over (2\pi)^3} A_0^{\mbox{in}} V^{-1}
 (\vert {\bf q} \vert )A_0^{\mbox{in}}    \; .
 \label{eq7}
\end{eqnarray}
Here $a_\mu^\pm $ and $A_0^{\mbox{in}}$ are fluctuating part of
the corresponding gauge fields. Note that the field $a_\mu^ -$
does not exist for unpolarized states and hence Eq.~(\ref{eq7})
reduces appropriately.
The polarization tensors $\Pi^{\mu\nu}_{\uparrow , \downarrow}$
have the following form,
\begin{eqnarray}
\Pi^{\mu \nu}_{ \uparrow , \downarrow } &=&
\Pi_0^{\uparrow , \downarrow } (\omega\, ,\, {\bf q}^2)
(q^2g^{\mu\nu} -q^\mu q^\nu )+ \left( \Pi_2^{\uparrow ,
\downarrow}-\Pi_0^{\uparrow , \downarrow} \right) (\omega \, ,
\, {\bf q}^2)  \nonumber \\
& & \times \left( {\bf q}^2 \delta^{ij} -q^{i}q^{j} \right)
\delta^{\mu i}\delta^{\nu j} +i\Pi_1^{\uparrow , \downarrow}
(\omega \, ,\, {\bf q}^2) \epsilon^{\mu\nu \lambda} q_ \lambda
\; . \label{eq8}
\end{eqnarray}
Integrating out  all the internal gauge fields,
the effective action for the external probes turns out to be
\begin{equation}
S_{eff} \left[ A_\mu^\uparrow ,A_\mu^\downarrow \right] = {1
\over 2} \int {d^3 q \over (2\pi)^3} A_\mu^r (q)
K^{\mu\nu}_{rr^\prime} (\omega \, ,\, {\bf q}^2)
A_\nu^{r^\prime} (-q) \; ,
\label{eq9}
\end{equation}
where the indices $r,r^\prime = \uparrow , \downarrow $.
$K^{\mu\nu}_{rr^\prime}$ measures linear response of the system
to weak external probes.
Recall that $K^{00}_{\uparrow
\uparrow}$, $K^{00}_{\uparrow \downarrow}$, $K^{00}_{\downarrow
\uparrow}$ and $K^{00}_{\downarrow \downarrow}$ represent the
density-density correlations among spin up-up, up-down, down-up
and down-down species of the particles respectively.
These are given by
\begin{mathletters}
\label{eq10}
\begin{eqnarray}
K^{00}_{\uparrow  \uparrow} &=& {{\bf q}^2 \over \Pi_0^\uparrow
+\Pi_0^\downarrow } \left[ \Pi_0^\uparrow \Pi_0^\downarrow
-{\left( \Pi_0^\uparrow \Pi_1^\downarrow -\Pi_0^\downarrow
\Pi_1^\uparrow +\Pi_0^\uparrow \theta_+ \right)^2 \over
{\cal D}(\omega \, , \, {\bf q})} \right]  \; , \\
K^{00}_{\downarrow  \downarrow} &=& {{\bf q}^2 \over \Pi_0^\uparrow
+\Pi_0^\downarrow } \left[ \Pi_0^\uparrow \Pi_0^\downarrow
-{\left( \Pi_1^\uparrow\Pi_0^\downarrow -\Pi_1^\downarrow
\Pi_0^\uparrow +\Pi_0^\downarrow \theta_+ \right)^2 \over
{\cal D}(\omega \, , \, {\bf q})} \right]  \; , \\
K^{00}_{\uparrow  \downarrow}= K^{00}_{\downarrow  \uparrow}
& =& -{{\bf q}^2 \over \Pi_0^\uparrow
+\Pi_0^\downarrow } \left[ \Pi_0^\uparrow \Pi_0^\downarrow
+{\left( \Pi_0^\uparrow\Pi_1^\downarrow -\Pi_0^\downarrow
\Pi_1^\uparrow +\Pi_0^\uparrow \theta_+ \right)
\left( \Pi_1^\uparrow\Pi_0^\downarrow -\Pi_1^\downarrow
\Pi_0^\uparrow +\Pi_0^\downarrow \theta_+ \right) \over
{\cal D}(\omega \, , \, {\bf q})} \right]  \; ,
\end{eqnarray}
\end{mathletters}
with
\begin{eqnarray}
{\cal D}(\omega \, , \, {\bf q}) &=&
\left( \Pi_0^\uparrow + \Pi_0^\downarrow \right)^2 \omega^2 -
\left( \Pi_1^\uparrow + \Pi_1^\downarrow +\theta_+ \right)^2 \nonumber
\\ & & - \left( \Pi_0^\uparrow + \Pi_0^\downarrow \right)
\left( \Pi_2^\uparrow + \Pi_2^\downarrow \right){\bf q}^2 -
\left( \Pi_0^\uparrow + \Pi_0^\downarrow \right) \theta_+^2
V({\bf q}^2){\bf q}^2  \; .
\label{eq11}
\end{eqnarray}

\section{Results and Experimental Consequences}

\subsection{Correlations}

The charge density correlation can now be obtained as
\begin{equation}
K^{00}(\omega \, ,\, {\bf q}^2) \equiv \sum_{r,r^\prime}
K^{00}_{rr^\prime}(\omega \, ,\, {\bf q}^2)=- {\bf q}^2 { \left(
\Pi_0^\uparrow +\Pi_0^\downarrow \right) \theta_+^2 \over {\cal
D} (\omega \, ,\, {\bf q}^2) } \; .
\label{eq12}
\end{equation}
On the other hand the spin density correlation is given by
\begin{equation}
\Sigma (\omega \, , \, {\bf q}^2) = \sum_{r,r^\prime} \left[
K^{00}_{rr^\prime}\delta_{rr^\prime} - K^{00}_{rr^\prime}
(1-\delta_{rr^\prime}) \right]
\label{eq13}
\end{equation}
For unpolarized states, $\Pi_0^\uparrow =\Pi_0^\downarrow \equiv
\Pi_0$.
Thus $\Sigma$ gets the simpler form,
\begin{equation}
\Sigma_{\rm unp} (\omega \, ,\, {\bf q}^2 )= \Pi_0
(\omega \, ,\, {\bf q}^2) {\bf q}^2 \; .
\label{eq14}
\end{equation}

Note at the outset that
the charge density excitations (CDE) will be very different from
spin density excitations (SDE) (spin-zero excitations)
especially for unpolarized states,
because, CDE are determined by the poles of $K^{00} (\omega \,
,\, {\bf q}^2)$, while SDE are determined by the poles of
$\Pi_0$. The collective CDE and SDE will be discussed
below. We note that the leading order term in ${\bf q}^2$ of $K^{00}$
saturates the f-sum rule \cite{lopez2}.

The above results
are valid in the thermodynamic limit.
For as Lopez and Fradkin \cite{lopez2} have argued in a similar
case, we note that we have
evaluated the effective action by neglecting higher order response
functions, viz, the correlations of three or more currents or
densities. These higher order correlations are of higher order in
${\bf q}^2$ compared to the quadratic term in Eqs.~(\ref{eq10},
\ref{eq12} -- \ref{eq14}).
These higher order
terms would not be negligible for a finite system since the minimum
allowed value of the momentum is then determined by the linear size
of the system $L$, i.e., $\vert {\bf q}\vert >1/L$.
On the other hand, in the
thermodynamic limit, $L \rightarrow  \infty $ and the minimum
allowed value of $\vert {\bf q}\vert $ goes to zero. Therefore
one is allowed to keep only the quadratic term in effective
action and neglect the higher order corrections for an
infinite system.

In the limit of low ${\bf q}^2$, CDC and SDC are respectively
given by
\begin{mathletters}
\label{eq15}
\begin{eqnarray}
K^{00}(\omega \, ,\, {\bf q}^2) &=& -\left( {e^2 \rho \over
m^\ast} \right) {1 \over \omega^2 -\omega_c^2}{\bf q}^2 +{\cal
O} (({\bf q}^2)^2)    \; , \\
\Sigma (\omega \, , \, {\bf q}^2) &=& -\left( {e^2\rho \over
m^\ast}\right) \left[
{(p_ \uparrow -p_ \downarrow )^2\over (p_ \uparrow + p_
\downarrow )^2}{1\over \omega^2 -\omega_c^2} +
{p_ \uparrow p_ \downarrow \over p_
\uparrow +p_ \downarrow }{1\over \omega^2 -\bar{\omega}_c^2}
\right] {\bf q}^2 + {\cal O}(({\bf q}^2)^2) \; .
\end{eqnarray}
\end{mathletters}
We see from Eq.~(\ref{eq15}) that
CDC preserves the Kohn mode \cite{kohn} of excitation. On the
other hand, SDC shows a new mode of excitation at
$\bar{\omega}_c$ apart from the actual cyclotron energy $\omega_c$.
Interestingly, in the case of unpolarized QHS for which $p_
\uparrow = p_ \downarrow $, {\em only} the mode at
$\bar{\omega}_c$ survives. This, in fact, gives the
measure of energy scale for CF.

\subsection{Spin Transition}

At $\omega =0$, SDC (\ref{eq15}) can be written as
\begin{equation}
\Sigma (0,{\bf q}^2) ={\bf q}^2 \left( {e^2 m^\ast \over 4\pi^2
\rho }\right) \left[
{ (p_ \uparrow
-p_ \downarrow )^2 \over (p_ \uparrow +p_ \downarrow )^2}\nu^2
 + p_ \uparrow p_ \downarrow      \right] \; .
\label{eq16}
\end{equation}
We see that $\Sigma (0,\, {\bf q}^2)$ given by the above
expression plays an important role in the spin transitions.
Eisenstein et al \cite{eisen} and Engel et al \cite{engel} have
observed spin transitions in QHS with filling fractions $\nu
=2/3$ and $3/5$. By the increase of Zeeman energy, QHS at $\nu
=2/3$ $(p_ \uparrow =p_ \downarrow =-1,\, s=1)$ and $\nu =3/5$
$(p_ \uparrow =-2,\, p_ \downarrow =-1,\, s=1)$ undergo a spin
transition from their respective phase of no polarization and
partial polarization to fully polarized phase $(p_ \downarrow
=0)$.
In this context, we note that the effective number of LL acts as an
order parameter in spin transition. Indeed,
the ratio of the values of $\Sigma$
between the unpolarized and fully polarized phases is given by
$\Sigma_{\rm unp}/\Sigma_{\rm p} =p_ \uparrow ^2 /\nu^2$. Therefore,
the ratio of $\Sigma (0, \, {\bf q}^2)$ in unpolarized and fully
polarized phase would determine $p_\uparrow (= p_\downarrow )$
in the unpolarized phase unambiguously; the ratio does
not depend on other parameters such as $m^\ast $ which has
complicated dependence on the magnetic field
\cite{lead,du2,mano}. Similarly, the ratio of $\Sigma (0, \,
{\bf q }^2 )$ in partially polarized and fully polarized would
also determine $p_\uparrow $ and $p_\downarrow $ in partially
polarized phase unambiguously. The
order parameter shows a discontinuity in the spin transitions.

\subsection{Neutron Scattering}

In the standard neutron scattering experiment \cite{mahan},
(in this case, the scattering is
in the plane of the sample), the differential
scattering cross section is given by
\begin{equation}
{d\sigma \over d\Omega }\propto {k_f \over k_i}\left[ S_c (q)
+{\sigma_\Sigma \over \sigma_c}S_\Sigma (q) \right] \; ,
\label{eq17}
\end{equation}
where ${\bf k}_i$ and ${\bf k}_f$ ar the momentum of the
incident and scattered neutrons, ${\bf q}={\bf k}_f -{\bf k}_i$
is the momentum transfer. $S_c (q)$ and $S_\Sigma (q)$ are
static charge and spin structure factors which are frequency
integrated imaginary part the corresponding correlation functions.
These are evaluated in this case, from Eq.~(\ref{eq15}), as
\begin{mathletters}
\label{eq18}
\begin{eqnarray}
S_c (q) &=& {\bf q}^2 \left( {e^2 \over 2}\right) \nu \; , \\
S_\Sigma (q) &=& {\bf q}^2 \left( {e^2 \over 2}\right) \left[
\left\vert {p_ \uparrow p_ \downarrow \over p_ \uparrow + p_
\downarrow }\right\vert + {(p_ \uparrow - p_ \downarrow )^2
\over (p_ \uparrow + p_ \downarrow )^2}\nu \right] \; .
\end{eqnarray}
\end{mathletters}
Note that unlike the parent expressions in Eq.~(\ref{eq15}), the
above expressions are free from the dependence on $m^\ast $
which by now is known to possess a dependence on the magnetic
field \cite{lead,du2,mano}.
In the unpolarized phase, $S_\Sigma (q)\propto p_ \uparrow $.
$S_c (q)$ is proportional to $\nu$ irrespective of the phase.
(In the fully polarized phase, $S_c (q)=S_\Sigma (q)$).
In Eq.~(\ref{eq17}),
$\sigma_\Sigma /\sigma_c $ is the ratio of the spin  and charge
dependent total cross sections. One can determine $S_c$ and
$S_\Sigma$ in unpolarized or partially polarized phases by two
different ways --- (i) By the measurement of cross section in
fully polarized phase, one will be able to extract $S_c (q)$
since cross section is proportional to $S_c (q)$. It is same in
all phases. The same experiment in unpolarized or partially
polarized phase, whichever is the relevant, has to be performed
to know $S_\Sigma (q)$. (ii) X-ray
scattering experiment will measure $S_c (q)$ and then neutron
scattering would determine $S_\Sigma (q)$ with the knowledge of
$S_c (q)$. Particularly in the unpolarized phase, $S_\Sigma (q)$
determines the composite fermion parameter $p_ \uparrow $.
In summary, neutron scattering experiment provides a direct
unambiguous test of CF. The accuracy of Eq.~(\ref{eq18})
lies on the region of small angle scattering as it is valid
only for low ${\bf q}^2$.

\subsection{Excitations and Raman Scattering}

We now determine the collective modes of CDE and SDE with respective
spectral weights. We discuss the excitations for both
unpolarized and fully polarized phases of the quantum Hall
state with $\nu
=2/3$ only in detail, as the state is observed in both the phases
\cite{eisen,engel}, and partly for simplicity. The calculation
for other states will follow a similar treatment. We use the
same procedure as Lopez and Fradkin \cite{lopez2} who have worked
out for fully polarized QHS. It should be possible to
observe the modes by polarized and depolarized resonant Raman
scattering. In this context, we note that
in inelastic light scattering experiments, the
magnetoplasmon modes of IQHE
and FQHE state at $\nu =1/3$ have been observed
\cite{pinc1,pinc2}.

We first consider CDE for fully polarized phase of $\nu =2/3$
$(p_\uparrow =-2,\, p_\downarrow =0, \, s=1)$ state. The modes
are determined from the poles of $K^{00}$. We look for the solutions
of the form \cite{lopez2} $\omega^2 = (k\bar{\omega}_c)^2 +\beta
(\bar{{\bf q}}^2)^\gamma $, where $\bar{{\bf q}}^2 ={\bf q}^2l_0^2/2$
with $l_0=(e\bar{B})^{-1/2}$ being the effective magnetic length,
$\beta$ and $\gamma$ are two constants to be determined for the
corresponding mode characterized by $k$ (an integer). The values of $k$
runs from 1 to 3. We find there are two modes for $k=2$ whose
dispersion relations are given by
\begin{equation}
\omega_{2\pm}^2 = (2\bar{\omega}_c)^2 +
\beta_\pm \bar{{\bf q}}^2
\label{eq19}
\end{equation}
with the corresponding residues in $K^{00}$ being
\begin{equation}
{\rm Res}(K^{00})\vert_{\omega_{2\pm}} = \mp \frac{1}{\pi
\bar{\omega}_c} \frac{\beta_\pm}{\beta_+ -\beta_-}\left[
\beta_\pm -3\bar{\omega}_c^2 \right] {\bf q}^2 \bar{{\bf q}}^2 \;
, \label{eq19a}
\end{equation}
where
\begin{equation}
\beta_\pm = {\bar{\omega}_c^2 \over
10} \left( 180 \pm \sqrt{(180)^2-15360} \right) \; .
\end{equation}
We do not find any mode whose zero momentum gap is at
$\bar{\omega}_c$. On the other hand, there are two modes at
$\omega_c$ (for ${\bf q}^2 =0$) with the dispersion relations
\begin{equation}
\omega_\pm^2 = \omega_c^2 -\bar{\omega}_c^2 \left[
\left( 14 +
{2m^\ast V(0) \over 2\pi}\right) \mp \sqrt{\left( 14+{2m^\ast
V(0)\over 2\pi }\right)^2 +2700 }
\right] \bar{{\bf q}}^2
\label{eq20}
\end{equation}
provided the interaction potential $V(q)$ is a regular function
at ${\bf q}^2 =0$. The residues in $K^{00}$ for the modes are
proportional to ${\bf q}^2$.
The residues are given by
\begin{equation}
{\rm Res}(K^{00})\vert_{\omega_\pm} = \pm {\bf q}^2 \omega_c
\frac{\nu}{8\pi}
\frac{\left( 14 +
{2m^\ast V(0) \over 2\pi}\right) \pm
\sqrt{\left( 14+{2m^\ast
V(0)\over 2\pi }\right)^2 +2700 } }{
\sqrt{\left( 14+{2m^\ast
V(0)\over 2\pi }\right)^2 +2700 }} \; .
\end{equation}
These modes have higher spectral
weights compared to the modes $\omega_{2\pm}$.

In the unpolarized phase of $2/3\, (p_ \uparrow =p_ \downarrow
=-1, \, s=1)$ state, the CDE modes for $k=2$ are given by
\begin{equation}
\omega_{2\pm}^2 = (2\bar{\omega}_c)^2 +
\alpha_\pm \bar{{\bf q}}^2
\label{eq21}
\end{equation}
with the residues in $K^{00}$ are proportional to ${\bf q}^4$
and they are given by
\begin{equation}
{\rm Res}(K^{00})\vert_{\omega_{2\pm}} = \mp \frac{1}{\pi
\bar{\omega}_c} \frac{\alpha_\pm}{\alpha_+ -\alpha_-}\left[
\alpha_\pm -3\bar{\omega}_c^2 \right] {\bf q}^2 \bar{{\bf q}}^2 \;
, \label{eq21a}
\end{equation}
where
\begin{equation}
\alpha_\pm = {\bar{\omega}_c^2 \over
10} \left( 48 \pm \sqrt{(48)^2-1920} \right) \; ,
\end{equation}
The other two modes for which the zero momentum gaps are at
$\omega_c$ follow
\begin{equation}
\omega_\pm^2 = \omega_c^2 -{\bar{\omega}_c^2 \over 20}
\left[
\left( 206 +20
{2m^\ast V(0) \over 2\pi}\right) \mp \sqrt{\left( 206+20{2m^\ast
V(0)\over 2\pi }\right)^2 +61 \times (120)^2 }
\right] \bar{{\bf q}}^2
\label{eq22}
\end{equation}
with the corresponding spectral weights are proportional to
${\bf q}^2$.
The residues of $K^{00}$ corresponding to these modes are given
by
\begin{equation}
{\rm Res}(K^{00})\vert_{\omega_\pm} = \pm {\bf q}^2 \omega_c
\frac{\nu}{8\pi}
\frac{ \left( 206 +20
{2m^\ast V(0) \over 2\pi}\right) \pm
\sqrt{\left( 206+20{2m^\ast
V(0)\over 2\pi }\right)^2 +61 \times (120)^2 } }{
\sqrt{\left( 206+20{2m^\ast
V(0)\over 2\pi }\right)^2 +61 \times (120)^2 } } \; .
\end{equation}
Similar to the fully polarized phase, no mode
exists for CDE at $\bar{\omega}_c$ (for ${\bf q}^2$).

We now determine SDE in the unpolarized phase
from the poles of $\Sigma_{\rm unp} (\omega \, ,\,{\bf q}^2)$
in Eq.(\ref{eq14}). Interestingly, SDE are at $\omega_k =
k\bar{\omega}_c$ ($k$ an integer) which do not disperse with
$\vert {\bf q}\vert$. Note that, unlike the CDE, SDE have a mode
at $\omega = \bar{\omega}_c$. The residue in $\Sigma$ for the
mode $\omega =\bar{\omega}_c$ is ${\rm Res}(\Sigma)=
\omega_c {\nu \over 2 \pi}{\bf q}^2$. The
spectral weights corresponding to other modes are proportional
to ${\bf q}^{2k}$.

We report here that for unpolarized QHS, the SDE have only one
dispersionless mode at $\omega = \bar{\omega}_c$.
All the other modes $\omega_k$ disperse with $\vert {\bf
q}\vert$ from the zero momentum value $k\bar{\omega}_c$.
The residue in $\Sigma$ for the
mode $\omega = \bar{\omega}_c$ is proportional to ${\bf q}^2$
and for all other dispersed modes $\omega_k$ $(k\neq 1)$, they
are down by a factor ${\bf q}^{2(k-1)}$. Therefore in the
unpolarized and partially polarized phase, unlike in the fully
polarized phase, SDE are very different from CDE.
Similarly for IQHE states, as have been obtained by Kallin and
Halperin \cite{kallin}, CDE and SDE are same for fully polarized
states, but they differ for partially polarized and unpolarized
states. Longo and Kallin \cite{longo} have studied
spin-flip and spin-wave excitations (which we do not consider
here) recently.

By polarized and depolarized Raman scattering experiments, the
modes of CDE and SDE can respectively be found out. The Raman
intensity $I(\omega)$ is proportional to the imaginary part of
the corresponding correlation functions \cite{klein}.

In the limit ${\bf q}^2l_0^2 \ll 1$, most of the weight of CDC
is in the cyclotron modes i.e., at $\omega_\pm$ (\ref{eq20} and
\ref{eq22}) for both unpolarized and fully polarized phases. The
accumulated contributions of these modes, in fact, saturate the
f-sum rule. The modes are degenerate in the limit ${\bf q}^2
\rightarrow  0$. The relative intensity for these modes
is given by
$I(\omega_+)/I(\omega_-) \sim 1$.
The splitting between the two modes $\Delta\omega^2 =\omega_+^2
-\omega_-^2$ is proportional to ${\bf q}^2$. The pole in CDC for
the excitation frequencies $\omega_{2\pm}$ may be read
off from Eqs.~(\ref{eq19}) and
(\ref{eq21}). Thus the intensities corresponding to these modes
$\omega_{2\pm}$ will be suppressed by a factor of ${\bf q}^2$
than the same for $\omega_\pm$ modes.

The situation for SDE in fully polarized and unpolarized phases
are very different. Depolarized Raman scattering experiment in
{\it fully polarized} phase creates a spectra very
similar to the one in the
polarized Raman scattering experiment because CDE and SDE are
same in this phase. On the other hand, in depolarized Raman
scattering in the {\it unpolarized} phase, the highest intensity will
be observed for the mode which is {\em exactly} at
$\bar{\omega}_c$. The intensity corresponding to the next higher
mode at $2\bar{\omega}_c$ is suppressed by a factor of $\bar{{\bf
q}}^2$ than the mode at $\bar{\omega}_c$. Similarly the
intensity for other modes are further down by factors of
$\bar{{\bf q}}^2$ compared to the previous lower mode. Although
we have discussed depolarized Raman spectra only for $\nu =2/3$
state in unpolarized phase, it is easy to check that the
characteristics of the spectra will be similar for all other
unpolarized QHS. Indeed, in all unpolarized and partially
polarized QHS, depolarized Raman spectra have the highest
intensity corresponding to the frequency $\omega
=\bar{\omega}_c$, the effective cyclotron frequency for
CF. In other words, the total effective number
of LL $(p_ \uparrow +p_ \downarrow )$ filled by CF
can be determined by the depolarized
Raman scattering experiments in the unpolarized or partially
polarized phases of relevant QHS \cite{model} as $\bar{\omega}_c
= (2\pi\rho /m^\ast) (1/(p_ \uparrow +p_ \downarrow ))$.
This determination will become exact if the effective mass
$m^\ast $ is determined independently. In any case,
$\bar{\omega}_c$ as the relevant scale would again be established
by this experiment. Importantly, note that this mode at
$\bar{\omega}_c$ is ${\bf q}^2 $ independent.

We remark that
polarized Raman scattering in any of the quantum Hall phases and
depolarized Raman scattering in fully polarized phase
measure the actual filling fraction $\nu$. On the other hand
in unpolarized and partially polarized phase, depolarized Raman
scattering measures the filling fraction of CF.

\section{CONCLUSION}

In summary, we state the most important results. Spin density
correlation (\ref{eq16}) represents an order parameter in the
spin transitions from unpolarized or partially polarized phases
to the fully polarized phase of the relevant quantum Hall states
\cite{model} as the Zeeman energy is increased. Spin density
correlation shows an undispersed mode at $\bar{\omega}_c$ in the
unpolarized and partially polarized phases. The spin density
excitations in these phases are very different from charge
density excitations. Neutron scattering and depolarized Raman
scattering experiments would directly determine one of the
composite fermion parameters, viz, the effective number of
filled Landau levels by CF. The other parameter
may be find out from the knowledge of the former.

Finally, Lopez and Fradkin \cite{lopez4} have studied recently
the bilayered QHS employing a similar model as ours. However,
the two models leads to certain different physical consequences.
(For detailed comparision between the two models, see
Ref.~\cite{wfn}). It might be of interest to examine whether there is
some experimental procedure that can determine the composite fermion
parameters in bilayered systems as well.


\begin{references}
\bibitem{fnote1} However, this apparent
       discrepency has been clarified later by Halperin, Lee and Read
       \cite{hlr} who have shown that the quasi-particle creation
       operator in the state $\nu = p/(2p+1)$
       creates a net charge of magnitude $-e/(2p+1)$ locally and
       pushes the missing charge $-2pe/(2p+1)$ to the the
	 boundary of the system as the addition of two flux quanta
	 creates a local charge deficit of $-2pe/(2p+1)$.
       The charges at the boundary get cancelled by the creation
	 of a quasi-particle at one point and the creation of a
	 quasi-hole at another point.
\end{references}
\end{document}